\documentclass[conference]{IEEEtran}
\IEEEoverridecommandlockouts
\def\e{{\mathbf e}}
\def\X{{\mathbf X}}
\def\Y{{\mathbf Y}}
\def\Z{{\mathbf Z}}
\def\L{{\mathcal L}}
\def\sg{\mathrm{sg}}
\def\vq{\mathrm{vq}}
\def\ce{\mathrm{ce}}
\def\se{\mathrm{se}}
\def\emb{\mathrm{emb}}
\def\R{\mathbb{R}}

\usepackage{amsmath,amssymb,amsfonts}
\usepackage{algorithmic}
\usepackage{graphicx}
\usepackage{textcomp}
\usepackage{xcolor}
\usepackage{bbding}
\usepackage{comment}
\usepackage[
backend=biber,
style=ieee,
citestyle=numeric-comp,
maxbibnames=3,
maxcitenames=3,
doi=false,isbn=false,url=false,eprint=false
]{biblatex}
\def\BibTeX{{\rm B\kern-.05em{\sc i\kern-.025em b}\kern-.08em
    T\kern-.1667em\lower.7ex\hbox{E}\kern-.125emX}}
\begin{document}

\title{Causal Speech Enhancement with \\Predicting Semantics based on \\Quantized Self-supervised Learning Features
\thanks{This work is supported by Research Grant S of the Tateishi Science and Technology Foundation.}
}

\author{\IEEEauthorblockN{Emiru Tsunoo, Yuki Saito, Wataru Nakata, Hiroshi Saruwatari}
\IEEEauthorblockA{\textit{The University of Tokyo}, Japan \\
\texttt{tsunoo@g.ecc.u-tokyo.ac.jp}}
}

% \author{\IEEEauthorblockN{Emiru Tsunoo}
% \IEEEauthorblockA{\textit{University of Tokyo}, Japan \\
% tsunoo@g.ecc.u-tokyo.ac.jp}
% \and
% \IEEEauthorblockN{Wataru Nakata}
% \IEEEauthorblockA{\textit{University of Tokyo}, Japan
% \\
% nakata-wataru855@g.ecc.u-tokyo.ac.jp
% }
% \and
% \IEEEauthorblockN{Yuki Saito}
% \IEEEauthorblockA{\textit{University of Tokyo}, Japan
% \\
% yuuki\_saito@ipc.i.u-tokyo.ac.jp
% }
% \and
% \IEEEauthorblockN{Hiroshi Saruwatari}
% \IEEEauthorblockA{\textit{University of Tokyo}, Japan
% \\
% hiroshi\_saruwatari@ipc.i.u-tokyo.ac.jp
% }
% }

\maketitle

\begin{abstract}
Real-time speech enhancement (SE) is essential to online speech communication.
Causal SE models use only the previous context while predicting future information, such as phoneme continuation, may help performing causal SE.
The phonetic information is often represented by quantizing latent features of self-supervised learning (SSL) models.
This work is the first to incorporate SSL features with causality into an SE model.
The causal SSL features are encoded and combined with spectrogram features using feature-wise linear modulation to estimate a mask for enhancing the noisy input speech.
Simultaneously, we quantize the causal SSL features using vector quantization to represent phonetic characteristics as semantic tokens. The model not only encodes SSL features but also predicts the future semantic tokens in multi-task learning (MTL).
The experimental results using VoiceBank + DEMAND dataset show that our proposed method achieves 2.88 in PESQ, especially with semantic prediction MTL, in which we confirm that the semantic prediction played an important role in causal SE.

\end{abstract}


\section{Introduction}
Speech is an efficient medium to communicate with other people or with voice agents via speech recognition.
However, there is often interference such as background noise and reverberations, which adversely affect the online communications and speech-recognition-based human-computer interactions.
Therefore, in such a use case, it is essential to apply speech enhancement (SE) in real-time using a causal model.

Deep-learning models for SE have been widely studied \cite{lu13_interspeech, xu2013experimental, sun2017multiple,tan2018convolutional}.
%The models were further improved by introducing convolution and recurrent layers and Deep Complex U-Net (DCUNET), which achieved best performance in Deep Noise Suppression challenge 2020 \cite{hu20g_interspeech}.
SE models were further improved by introducing convolution and recurrent layers \cite{tan2018convolutional}, Deep Complex U-Net (DCUNET) \cite{choi2018phase} inspired by the success of image processing \cite{ronneberger2015u}, and their combination \cite{hu20g_interspeech}, which achieved best performance in Deep Noise Suppression challenge 2020 \cite{reddy20_interspeech}.
% Tan {\it et al.} combined convolution and recurrent layers for speech enhancement \cite{tan2018convolutional}. 
% Choi {\it et al.} \cite{choi2018phase} proposed Deep Complex U-Net (DCUNET) inspired by the success of image processing \cite{ronneberger2015u}.
% The DCUNET architecture was further combined with LSTM layers \cite{hu20g_interspeech} and achieved best performance in Deep Noise Suppression challenge 2020 \cite{reddy20_interspeech}.
MP-SENet \cite{lu23e_interspeech} further pushed up state-of-the-art results by directly denoising magnitude and phase in parallel.
% On the other hand, MFNet \cite{liu23c_interspeech} applied mapping instead of masking of the spectral features without phase information using U-Net-based architecture.
%However, these methods process in batch using the entire utterances.
Because SE is also important in real-time use-cases, some studies take into account causal systems.
Conv-TasNet directly models time-domain signals using convolution \cite{luo2019conv}.
The network architectures were further explored \cite{yuan2020time, valin20_interspeech,lv2021dccrn+, chen22c_interspeech} or applied adaptation using meta learning \cite{yu22_interspeech}.

% Self-supervised learning (SSL) uses a significant amount of unlabeled speech data, which is also useful for representing speech.
% Contrastive predictive coding \cite{oord2018representation}, wav2vec \cite{schneider19_interspeech}, HuBERT \cite{hsu2021hubert}, and WavLM \cite{chen2022wavlm} are known to enhance tasks such as speech recognition \cite{hsu2021hubert_icassp}, emotion recognition \cite{wang2021fine}, and speech translation \cite{nguyen20_interspeech}.
Self-supervised learning (SSL), such as WavLM \cite{chen2022wavlm}, uses a significant amount of unlabeled speech data, which is useful for representing speech.
Recent SE tasks also incorporate the SSL models \cite{wang2020self,zezario2020self,hung22_interspeech}.
SSL features are further quantized typically with $k$-means clustering to be used as semantic tokens, which are often highly correlated to phoneme states of speech \cite{lakhotia2021generative, sicherman2023analysing}.
In SELM \cite{wang_selm}, latent features of WavLM are quantized as semantic tokens, and the language model (LM) predicts the token sequence to re-synthesize enhanced speech.
However, less studies consider causal systems because SSL models are not causal by their nature.
For audio generation tasks, acoustic tokens are also used \cite{yang2023uniaudio,wang2024speechx}.

This work aims to exploit SSL models for SE with causality.
Even in the causal scenario, which has access only to the past context, predicting future information is found to be effective in incremental text-to-speech (TTS) \cite{saeki2021low} and streaming voice conversion (VC) \cite{ning24_interspeech}.
Inspired by them, we predict future phonetic information for causal SE.
We first encode the SSL features computed with causality, and then the encoded features are combined with the original noisy spectrogram features to predict a mask for SE.
Simultaneously, we quantize the causal SSL features using vector quantization (VQ), as semantic tokens. 
During training the encoder, we predict future semantic tokens using an LM, as in \cite{saeki2021low,ning24_interspeech}, in a multi-task learning (MTL) manner, so that the model becomes aware of phoneme continuation.
In experiments, we train the model with VoiceBank + DEMAND dataset \cite{veaux2013voice, thiemann2013diverse, valentini2016investigating} to see the effectiveness of incorporating semantic token prediction in the causal SE task.
We confirm that the semantic token prediction is particularly useful for causal SE.
Our contributions are as follows: 
\begin{enumerate}
    \item To the best of our knowledge, this is the first work to incorporate causal SSL features into an SE model.
    \item The training is in one-stage; the SSL features are simultaneously quantized using VQ as semantic tokens without a two-stage individual clustering process.
    \item We train an SSL feature encoder with future semantic token prediction as MTL, so that the model becomes aware of future phoneme information.
    \item Our methods achieved 2.88 in PESQ on the VoiceBank + DEMAND test set especially when we use semantic prediction MTL, which increased PESQ by 0.05.
\end{enumerate}

%\section{Conventional Speech Enhancement with \\Self-supervised learning representation}
\section{Conventional SE with SSL features}
%\subsection{Baseline system}
We adopt an SE mask estimation model described in \cite{hung22_interspeech} as the baseline.
The model combines SSL features with $\log 1p$ spectrogram features \cite{fu2020boosting} and estimate mask function to enhance speech components in the input signals.
Following \cite{hung22_interspeech}, the SSL features of each layer are averaged with weighted-sum.
The model output is scaled into $[0, 1]$ using the sigmoid function as a mask and multiplied by the input spectrogram.

Let $\X\in \R^{T\times F}$ be the $T$-frame input noisy spectrogram and its $\log 1p$ features be $\X'$ as
\begin{align}
\X' = \log(1+|\X|).    \label{eq:log1p}
\end{align}
Let $s_i(\X)$ be $i$-th layer output of an SSL model ($1 \leq i \leq I$), and $f$ be the mask estimation function.
With trainable weight $w_i$, the model estimates enhanced speech $\hat{\X}$ as follows.
\begin{align}
    s(\X) &= \sum_{i=1}^{I}w_i s_i(\X), \label{eq:ssl}\\
    \hat{\X}' &= \X' \odot \sigma(f(\X'\oplus s(\X))), \label{eq:mask}
\end{align}
where $\sigma$ is a sigmoid function, $\odot$ is element-wise multiplication, $\oplus$ is a concatenation operation, and $f$ is a combination of linear layers and bi-directional long-short time memory (LSTM) layers.
The SE model $f$ is trained by minimizing the L1 loss with clean speech $\log 1p$ spectrogram $\Y' \in \R^{T \times F}$, as
\begin{align}
    \L_{\se} &= ||\Y'-\hat{\X}'||_1^1. \label{eq:se_loss}
\end{align}
By this optimization, the SSL model can also be updated, and according to \cite{hung22_interspeech}, updating the parameters of only the Transformer layers in the SSL model was effective the most.
To reconstruct the waveform, the original noisy phase is applied.

\section{Modifications for Causal Models}

\subsection{Causal SSL}
Most of the SSL models, including WavLM \cite{chen2022wavlm}, are designed to process in batch.
Therefore, it is generally difficult to use the pretrained models in a causal scenario.
We approximate it by restricting to use only past frames and obtain each frame output, as in \cite{tsunoo2022run}, i.e., for time frame $t$, we take the last frame from $s(\X_{\leq t})$; thus the causal SSL features $c(\X)$ are represented as
\begin{align}
    c(\X) = \{s(\X_{\leq t})_{t}|1\leq t \leq T\}.
\end{align}
It is desirable to train an SSL model that is particularly designed for a causal system, which remains for our future work.

\subsection{Causal Transformer}
\label{ssec:trans}
For the mask estimation function $f$, the original model \cite{hung22_interspeech} uses bi-directional LSTM, which requires the entire utterance.
Although it can be replaced with uni-directional LSTM, we propose using the causal Transformers \cite{vaswani2017attention} in this work.
We design a mask so that the self-attention layers can only attend to past frames.
We compare the causal Transformer with LSTM in an ablation study in Sec.~\ref{ssec:ablation}.

\subsection{Feature-wise linear modulation (FiLM)}
\label{ssec:film}
FiLM \cite{perez2018film} is known to be an effective method to combine two modalities.
While the original method concatenates $\log 1p$ features with SSL features, we apply FiLM to combine two features.
The concatenation operation $\oplus$ in \eqref{eq:mask} is replaced by the FiLM manipulation, which is represented by $\otimes$ as
\begin{align}
    \X'\otimes c(\X) = \gamma(c(\X)) \odot \alpha(\X') +  \beta(c(\X)), \label{eq:oplus}
\end{align}
where $\alpha$, $\beta$, and $\gamma$ are linear functions.
FiLM is also compared with naive concatenation in Sec.~\ref{ssec:ablation}.

\section{Semantic Prediction in Causal System}
\label{sec:semantic}
\begin{figure}[t]
\includegraphics[width=1.0\columnwidth]{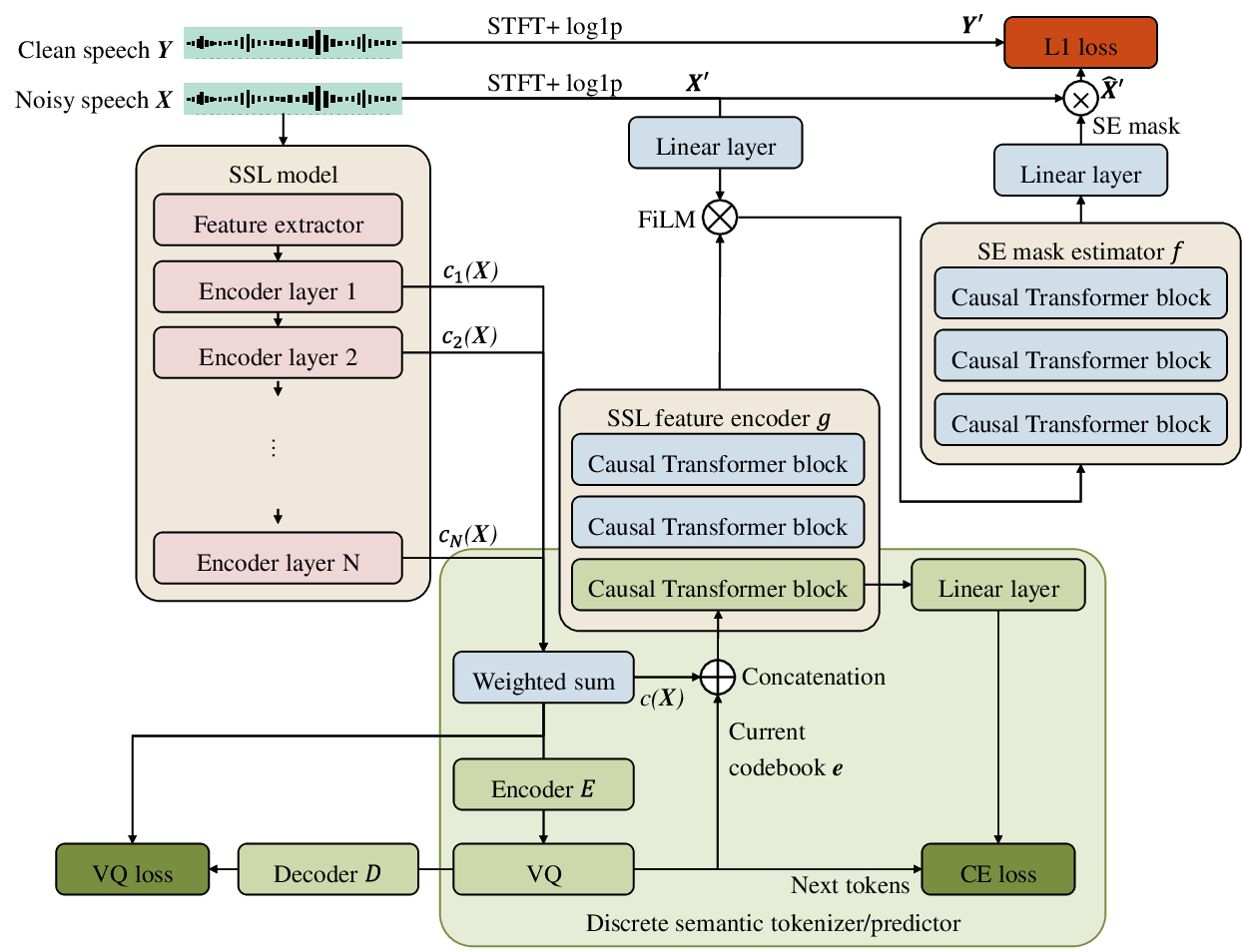}
\caption{SE model overview.}
\label{fig:overview}
\vspace{-0.3cm}
\end{figure}

In a causal scenario, the mask needs to be estimated using only the past information, while linguistic structures and semantics in future can be easily predicted.
The semantic prediction can be performed similarly to LMs predicting the next word from its context.
Inspired by the success of token prediction in incremental TTS \cite{saeki2021low} and streaming VC \cite{ning24_interspeech}, we further incorporate semantic prediction into the causal SE model.
We use an additional SSL feature encoder $g$, which not only encodes the causal SSL features but also predicts future semantics from the context.
The semantics are modeled as quantized tokens in \cite{wang_selm}.
Instead of independent clustering of the causal SSL features using the $k$-means algorithm, we adopt VQ as in \cite{huang2023repcodec}.
The quantized semantic tokens are predicted from their semantic context and the input causal SSL features.
The semantic prediction is trained in an MTL manner so that the SSL feature encoder $g$ becomes aware of semantic continuation.
The overview of the proposed SE method is presented in Fig.~\ref{fig:overview}.

\subsection{Discretized semantic representation with VQ}
SSL features are often quantized as semantic information, so that an autoregressive LM can easily predict the next discrete semantic token from the context \cite{wang_selm}.
As discussed in \cite{huang2023repcodec}, the discretization of SSL features can be done with either $k$-means or VQ, and they only differ in terms of optimization; thus we adopt VQ.
Unlike two-stage individual quantization, the VQ training is done simultaneously with SE model training so that the entire training is carried out in one-stage.

The causal SSL features with weighted-sum ($c(\X)$) are encoded by a linear layer as $E(c(\X))$.
Then the encoded features are quantized with codebooks by selecting the closest codebook $\e$ to the encoded output.
The codebook is considered to be a discretized semantic token.
Subsequently, the codebook is decoded by another linear layer as $D(\e)$ to reconstruct $c(\X)$.
We also use moving average technique \cite{razavi2019generating} and combine it to the commitment loss as VQ loss shown in lower left of Fig.~\ref{fig:overview}.
\begin{align}
    \L_{\vq} = ||c(\X)-D(\e)||_2^2 &+ ||\sg[E(c(\X))]-\e||_2^2 \nonumber \\&+ \xi || \sg[\e]-E(c(\X))||_2^2, \label{eq:vqloss}
\end{align}
where $\sg$ represents stop gradient and $\xi$ is a tunable parameter.
We use simple linear projection in both the encoder $E$ and decoder $D$ because we make the quantization perform similarly to $k$-means clustering of the causal SSL features $c(\X)$.
Note that the VQ training is carried out using the causal SSL features that are also updated with the trainable averaging weights in \eqref{eq:ssl}.
Thus, we use the quantized VQ codebooks as discrete semantic tokens, which are to be predicted in the model described in the following subsection.

\begin{figure}[t]
\includegraphics[width=0.9\columnwidth]{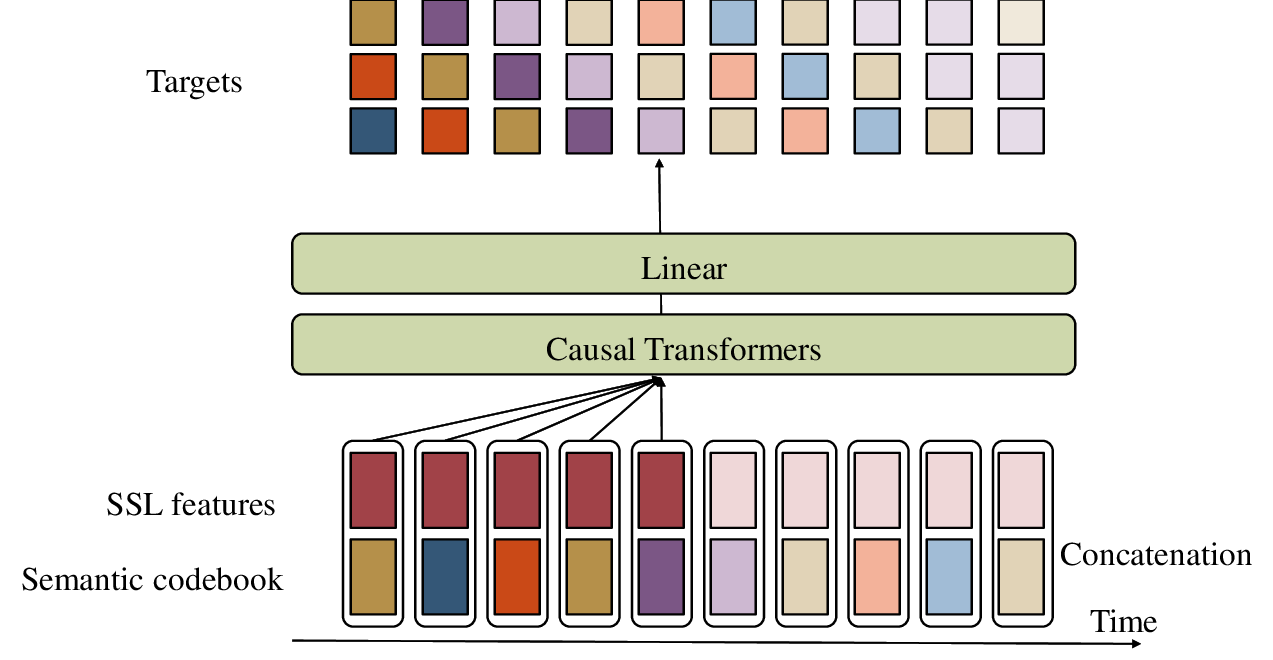}
\caption{Semantic token prediction model. The model predict next 3 tokens simultaneously in this example.}
\label{fig:prediction}
\vspace{-0.3cm}
\end{figure}

\subsection{MTL with semantic prediction}
\label{ssec:semantic}
During training, the SSL feature encoder $g$ is also trained with the semantic prediction task.
Causal SE uses only the past frames and future semantic prediction can help improve mask estimation, as in incremental TTS \cite{saeki2021low} and streaming VC \cite{ning24_interspeech}.
We add a branch to the SSL feature encoder $g$ and predict the next $N$ semantic tokens.

While the SSL feature encoder $g$ uses the causal SSL features $c(\X)$ as the input, additional discrete information can help prediction similarly to LMs using discrete context for prediction.
\subsubsection{Raw SSL features as input}
Let the input of the SSL feature encoder $g$ be $\Z$.
In the case that only the causal SSL features are used, the input is defined as
\begin{align}
    \Z = c(\X) \label{eq:raw_input}.
\end{align}
\subsubsection{Codebook index as additional input}
We can also use the codebook index of the current token as additional information; thus
\begin{align}
    \Z = c(\X) \oplus \emb(\hat{e}), \label{eq:embedding_input}
\end{align}
where $\emb$ is an embedding look-up table and $\hat{e}$ is the codebook index.
\subsubsection{Latent codebook vector as additional input}
Instead of using the index, the codebook vector of the current frame itself can be used, which is distributed in the encoded space of VQ:
\begin{align}
    \Z = c(\X) \oplus \e. \label{eq:codebook_input}
\end{align}

The procedure of semantic prediction using latent codebook vector \eqref{eq:codebook_input} is presented in Fig.~\ref{fig:prediction}.
%We concatenate SSL features and a VQ codebook $\e$ of the current semantic token as an input to the SSL feature encoder.
%Consequently, we added a branch to the encoder and predicted the next $N$ tokens.
Similar to the reduction factor in \cite{meng2024autoregressive}, the next $N$ tokens are simultaneously predicted.
Fig.~\ref{fig:prediction} shows an example of $N=3$.
%The SSL features $s(\X)$ and VQ codebook $\e$ are first concatenated as the input for the prediction model.
The model predicts the probability of the next $N$ tokens, as
\begin{align}
    p(\e_{t+1:t+N}|\Z_{\leq t}) = \prod_{n=1}^N p(\e_{t+n}|\Z_{\leq t}). \label{eq;prediction}
\end{align}
We minimize the cross-entropy loss of the predicted token and the next tokens:
\begin{align}
    \L_{\ce} = \sum_{t=1}^{T-N} \sum_{n=1}^{N} q(\e_{t+n}) \log p(\e_{t+n}|\Z_{\leq t}),
\end{align}
where $q(\e)$ is the reference distribution given by applying VQ to the future frames.

\subsection{Training loss}
Training of SE, VQ, and semantic prediction are carried out in parallel as MTL.
% As described in Sec.~\ref{ssec:semantic}, we incorporate the learned codebook into the SE mask estimation and 
SE in \eqref{eq:mask} becomes
\begin{align}
    \hat{\X}' &= \X' \odot \sigma(f(\X'\otimes g(\Z))). \label{eq:mask2}
\end{align}
%In addition to L1 loss of enhanced speech $\hat{\X}$ and the reference clean speech $\Y$ described in \cite{fu2020boosting}
In addition to the SE loss in \eqref{eq:se_loss}, we combine the aforementioned losses with tunable weights $\lambda$'s.
\begin{align}
    \L & = \lambda_\se \L_{\se} + \lambda_\vq \L_{\vq} + \lambda_\ce \L_{\ce} \label{eq:loss}
\end{align}
By minimizing \eqref{eq:loss}, the causal SE model with semantic prediction is trained.

\section{Experiments}
\subsection{Experimental setup}
\label{ssec:setup}
To evaluate efficacy of our proposed semantic-continuation-aware causal SE model, we conduct experiments using Voice Bank-DEMAND dataset \cite{veaux2013voice, thiemann2013diverse, valentini2016investigating}. 
We evaluated PESQ and STOI for speech quality and
intelligibility, respectively. 

Our model consisted of three-layer causal Transformers for both the SSL encoder $g$ and SE mask estimator $f$ in \eqref{eq:mask2}, which had 4 heads and the numbers of hidden units were 512 and 256, respectively.
We adopted WavLM base \cite{chen2022wavlm} as the SSL model.
The VQ codebook size was 1024 and the semantic predictor predicted next $N=5$ tokens in \eqref{eq;prediction} unless otherwise stated.
We set $\xi=0.1$ in \eqref{eq:vqloss} and $\{\lambda_\se,\lambda_\vq,\lambda_\ce\}=\{1,1,0.01\}$ in \eqref{eq:loss} to balance the losses.
The model was trained for 200 epochs with a learning rate of 0.0001 and the Adam optimizer.% \cite{diederik2014adam}.

\subsection{SE results}
We first compared the input features for our proposed SE model as discussed in Sec.~\ref{ssec:semantic}.
Table~\ref{tb:inputs} lists the results of using raw SSL features \eqref{eq:raw_input}, of adding codebook index \eqref{eq:embedding_input}, and of adding codebook latent vector \eqref{eq:codebook_input}.
CSIG, CBAK and COVL are also reported, which are commonly used to measure signal distortion, noise distortion, and overall quality evaluation, respectively \cite{hu2007evaluation}.
We observed that using codebook latent vector, which were also optimized not only with VQ loss $\L_{\vq}$ in \eqref{eq:vqloss} but also with SE loss $\L_{\se}$ in \eqref{eq:se_loss}, performed the best among the input features.
We used the latent vector as input for the following experiments.

Subsequently, we compared our proposed method to the other causal SE methods, which include ConvTasNet \cite{luo2019conv}, the temporal convolutional network (TCN) \cite{koyama2020exploring}, Transformer \cite{fu2020boosting}, the convolutional recurrent neural network (CRN), time--frequency smoothing based on LSTM \cite{yuan2020time}, DCCRN+ \cite{lv2021dccrn+}, and LFSFNet \cite{chen22c_interspeech}.
%We also compared among input features $\Z$ for the SSL feature encoder $g$ defined in \eqref{eq:raw_input}, \eqref{eq:embedding_input}, and \eqref{eq:codebook_input}.
The results are listed in Table~\ref{tb:results}.
The numbers of the conventional methods are based on the literature.
We observed that our proposed methods outperformed most of the other conventional methods, and comparable to state-of-the-art method that, unlike our model, estimates appropriate phase information.
When we excluded semantic prediction from our model by setting $\lambda_{\ce}=0$, the performance degraded down to 2.83 in PESQ from 2.88.
This clearly indicates that the proposed semantic prediction MTL contributes to significant improvements on SE performance.
%When we used learned codebooks as the additional input ($\Z$ in \eqref{eq:codebook_input}) achieved best performance with 2.88 of PESQ..

\subsection{Effect of the number of token prediction on SE}
We then changed the numbers of semantic token prediction done in each frame and investigated the effect of them on SE performance.
$N$ in \eqref{eq;prediction} was varied from 1 to 10 and we evaluated not only PESQ but also the accuracy of the token prediction.
The results are shown in Table~\ref{tb:pred_num}.
As we increased the number of prediction $N$, the accuracy decreased because it is generally difficult to predict far future information.
This accuracy drop adversely affect SE performance as PESQ decreased in the case of $N=8$ and $N=10$ comparing to the case of $N=5$.
Thus, we confirmed $N=5$ performed the best in the Voice Bank-DEMAND test set.

\begin{table}[t]
\scriptsize
\caption{Results of input feature combinations for semantic prediction}
\vspace{-0.3cm}
\begin{center}
\begin{tabular}{|c|c|c|c|c|c|}
\hline
\textbf{Input features} & \textbf{PESQ} & \textbf{CSIG} & \textbf{CBAK} & \textbf{COVL} & \textbf{STOI} \\
\hline\hline
{Only SSL \eqref{eq:raw_input}} & {2.80} & 4.22 & 3.35 & 3.51& {\bf 0.94} \\ % train_bsse21.sh
{SSL + codebook index \eqref{eq:embedding_input}} & {2.86} & {4.21} & 3.36 & 3.52 &  {\bf 0.94} \\ % train_bsse20.sh
{SSL + codebook vector \eqref{eq:codebook_input}} & {\bf 2.88} & {\bf 4.28} & {\bf 3.39} & {\bf 3.58}& {\bf 0.94} \\ %train_bsse14.sh
\hline
\end{tabular}
\label{tb:inputs}
\end{center}
\vspace{-0.5cm}
\end{table}

\begin{table}[t]
\scriptsize
\caption{Evaluation results of causal SE systems
on the VoiceBank-DEMAND dataset}
\vspace{-0.7cm}
\begin{center}
\begin{tabular}{|c|c|c|c|c|c|}
\hline
\textbf{SE methods} & \textbf{PESQ} & \textbf{CSIG} & \textbf{CBAK} & \textbf{COVL} & \textbf{STOI} \\
\hline\hline
ConvTasNet \cite{koyama2020exploring} & 2.53 & 3.95 & 3.08 & 3.23 & - \\
STFT-TCN \cite{koyama2020exploring} & 2.73 & 4.11 & 3.25 & 3.42 & - \\
Transformer \cite{fu2020boosting} & 2.69 & 4.07 & 3.03 & 3.38 & 0.93 \\
CRN-MSE \cite{tan18_interspeech} & 2.74 & 3.86 & 3.14 & 3.30 & 0.93 \\
TFSNN \cite{yuan2020time} & 2.79 & 4.17 & 3.27 & 3.49 & -\\
DCCRN+ \cite{lv2021dccrn+}& 2.84 & - & - & - & - \\
LFSFNet \cite{chen22c_interspeech}& {\bf 2.91} & - & - & - & -\\
% Causal WavLM SE & 2.82 & & & & {\bf 0.94} \\
\hline
{Causal SSL w/o Semantic Pred.} & {2.83} & 4.23 & 3.36 & 3.53 &  {\bf 0.95} \\ % train_bsse20.sh
{Causal SSL w/ Semantic Pred.} & {2.88} & {\bf 4.28} & {\bf 3.39} & {\bf 3.58}& {0.94} \\ %train_bsse14.sh
% N=1: train_bsse16.sh
% N=3: train_bsse17.sh
% N=8: train_bsse18.sh
% N=10: train_bsse19.sh
\hline
\end{tabular}
\label{tb:results}
\end{center}
\vspace{-0.7cm}
\end{table}

\begin{table}[t]
\scriptsize
\caption{Comparison of the number of prediction per frame}
\vspace{-0.3cm}
\begin{center}
\begin{tabular}{|c|c|c|c|c|c|}
\hline
\textbf{\# of prediction} & \textbf{$N=1$} & \textbf{$N=3$} & \textbf{$N=5$} & \textbf{$N=8$} & \textbf{$N=10$} \\
\hline\hline
{PESQ} & {2.81} & 2.85 & {\bf 2.88} & 2.83 & 2.82 \\ % train_bsse21.sh
{Accuracy} & {33.7\%} & 21.5\% & 17.0\% & 13.0\% & 10.3\% \\ % train_bsse20.sh
% N=1: train_bsse16.sh
% N=3: train_bsse17.sh
% N=8: train_bsse18.sh
% N=10: train_bsse19.sh
\hline
\end{tabular}
\label{tb:pred_num}
\end{center}
\vspace{-0.5cm}
\end{table}

\begin{table}[t]
\scriptsize
\caption{Ablation study of causal SE on the VoiceBank-DEMAND}
\vspace{-0.3cm}
\begin{center}
\begin{tabular}{|c|c|c|c|c|c|c|}
\hline
\textbf{ID}&\multicolumn{4}{|c|}{\textbf{Conditions}} & \textbf{PESQ} &\textbf{STOI} \\
&\textbf{SE model} & \textbf{WavLM} & \textbf{FiLM} & \textbf{Semantics} &&\\
\hline\hline
C1 & Causal Trans. & Causal & \checkmark & \checkmark & 2.88 & 0.94 \\ 
C2 & Causal Trans. & Causal & \checkmark & & 2.83 & 0.94 \\
C3 & Causal Trans. & Causal & & & 2.67 & 0.94 \\
S1 & Causal Trans.  & None &  & & 2.64 & 0.94 \\
S2 & LSTM & None & & & 2.58& 0.94\\
N1 & Trans. & Non-causal & \checkmark & \checkmark & 3.14 & 0.96 \\ % PF train_bsse11.sh / None train_bsse3.sh
N2 & Trans. & Non-causal & \checkmark & & 3.14 & 0.96 \\
N3 & Trans. & Non-causal &  & & 3.09 & 0.95\\
 & BLSTM \cite{hung22_interspeech} & Non-causal &  & & 3.08 & 0.95\\
\hline
\end{tabular}
\label{tb:ablation}
\end{center}
\vspace{-0.4cm}
\end{table}

\subsection{Ablation study of model architectures}
\label{ssec:ablation}
We conducted an ablation study to see how the each element in our model contributed the SE performance in both the causal cases and the non-causal cases.
In the non-causal systems, causal Transformers described in Sec.~\ref{ssec:trans} were replaced with an ordinary Transformers so that both the SE model and WavLM had access to the entire context.
PESQ and STOI in various conditions are shown in Table~\ref{tb:ablation}.
The reproduction of original SSL-based method \cite{hung22_interspeech} is also listed as a reference.

First, when we compared FiLM in Sec.\ref{ssec:film} with simple concatenation, FiLM resulted in higher PESQ values in both the causal (C2 vs C3) and non-causal scenarios (N2 vs N3).
Furthermore, when we used only $\log 1p$ spectrogram features (S1), we observed degradation from the C3 condition.
%when we excluded semantic prediction (C1 vs C2), which was key idea of this work, PESQ drops significantly.
In non-causal systems, semantic prediction did not contributed SE performance (N1 vs N2), which was reasonable because the future information was already included in the context for the non-causal systems.
%Secondly, 
Lastly, by replacing causal Transformer with uni-directional LSTM we can see that our modification described in Sec.~\ref{ssec:trans} contributed SE performance by 0.06 PESQ improvement (S1 vs S2).

\section{Conclusion}
We proposed the causal SE model that predicted semantics of the future content.
The causal SSL features were calculated frame-by-frame using only the past input, and combined with $\log 1p$ spectrogram features with FiLM.
Simultaneously, the causal SSL features were quantized by VQ as semantic tokens, and the future tokens were predicted by the branch of the SSL feature encoder.
Experiments using Voice Bank-DEMAND showed that our proposed future-semantic-aware SE model achieved 2.88 in PESQ. 
We observed that the semantic prediction played an important role in causal SE.

Future work includes training causal SSL models because we forcibly used the pre-trained SSL models for the causal SE.
We will also adopt state-of-the-art model architectures to estimate appropriate phase information for further SE performance.
In addition, the causal SE model can also use other discrete representations such as neural audio codec for future prediction.

%\bibliographystyle{IEEEbib}
%\bibliography{utokyobib}
\renewcommand*{\bibfont}{\fontsize{9}{11}\selectfont}
\printbibliography

\end{document}